\documentclass[conference,a4paper]{APSIPA2021}
\usepackage{multirow}
\usepackage{graphicx}
\usepackage{amsmath}
\usepackage[psamsfonts]{amssymb}
\usepackage{amsxtra}
\usepackage{threeparttable}
\usepackage{mathtools}
\usepackage{url}

\begin{document}

\title{Conditional Deep Hierarchical Variational Autoencoder for Voice Conversion}

\author{%
    \authorblockN{%
    Kei Akuzawa\authorrefmark{1}\authorrefmark{2} and
    Kotaro Onishi\authorrefmark{1}\authorrefmark{3} and
    Keisuke Takiguchi\authorrefmark{1} and
    Kohki Mametani\authorrefmark{1} and
    Koichiro Mori\authorrefmark{1}
    }
    \authorblockA{%
    \authorrefmark{1}
    DeNA Co., Ltd. Tokyo, Japan
    }

    \authorblockA{%
    \authorrefmark{2}
    University of Tokyo, Tokyo, Japan
    }

    \authorblockA{%
    \authorrefmark{3}
    The University of Electro-Communications, Tokyo, Japan
    }

    \authorblockA{%
    E-mail: akuzawa-kei@weblab.t.u-tokyo.ac.jp, \{kotaro.onishi, keisuke.takiguchi, koki.mametani, koichiro.mori\}@dena.com
    }
}

\maketitle
\thispagestyle{empty}

\begin{abstract}
Variational autoencoder-based voice conversion (VAE-VC) has the advantage of requiring only pairs of speeches and speaker labels for training.
Unlike the majority of the research in VAE-VC which focuses on utilizing auxiliary losses or discretizing latent variables, this paper investigates how an increasing model expressiveness has benefits and impacts on the VAE-VC.
Specifically, we first analyze VAE-VC from a rate-distortion perspective, and point out that model expressiveness is significant for VAE-VC because rate and distortion reflect similarity and naturalness of converted speeches.
Based on the analysis, we propose a novel VC method using a deep hierarchical VAE, which has high model expressiveness as well as having fast conversion speed thanks to its non-autoregressive decoder.
Also, our analysis reveals another problem that similarity can be degraded when the latent variable of VAEs has redundant information.
We address the problem by controlling the information contained in the latent variable using $\beta$-VAE objective.
In the experiment using VCTK corpus, the proposed method achieved mean opinion scores higher than 3.5 on both naturalness and similarity in inter-gender settings, 
which are higher than the scores of existing autoencoder-based VC methods.
\end{abstract}

\section{Introduction} \label{sec:intro}
Voice conversion (VC) \cite{stylianou1998vc} is a technique for modifying a speech from a source speaker to match the vocal characteristics of a target speaker while keeping its phonetic content.
VC can be employed in many practical applications such as speaking aids \cite{kain2007said} and entertainment (e.g., singing VC \cite{doi2012singing}).

Although continued research efforts have improved overall quality of VC, there are still problems that prohibits the technique from being used in production.
For example, many conventional approaches \cite{toda2007vc,tanaka2019seq2seq} rely on parallel corpora in which the linguistic contents of speech data of multiple speakers are aligned.
It is known that collecting such data is expensive even for one-to-one conversion, let alone more practical situation like many-to-one and many-to-many conversion.
Also, recent studies \cite{sun2016ppg,liu2020vcc} proposed utilizing pre-trained speech recognition and synthesis models for VC.
This approach, which is referred to as Phonetic PosteriorGram-VC (PPG-VC), has several advantages:
it does not require parallel corpora and can be applied to a many-to-one VC task.
However, pre-training the speech recognition model needs speech transcription.
Moreover, pre-training the speech synthesis model with expressive or noisy speech corpora remains difficult although some recent studies have tackled the problem \cite{hsu2018hierarchical,hsu2019disentangling}.

Compared to the methods above, Variational Autoencoder-based VC (VAE-VC) \cite{hsu2016vaevc} does not require parallel corpora nor text transcriptions.
Instead, it can train on pairs of speeches and speaker labels.
In addition, since VAE-VC learns by reconstructing speech rather than mapping from text (or PPG) to speech, it is possible in principle to incorporate speeches into training regardless of the quality of them.
However, the conversion quality of VAE-VC is still limited compared to PPG-VC as shown in voice conversion challenge (VCC) 2020 \cite{yi2020vcc}, even though many researches have improved the quality by utilizing auxiliary losses \cite{hsu2017vawgan,kameoka2018acvae,tobing2019non} or discretizing latent variables \cite{van2017neural,wu2020one}.

Unlike the existing research, this paper investigates how an increasing model expressiveness has benefits and impacts on the VAE-VC.
Specifically, we first conduct a rate-distortion (RD) analysis and show that the model expressiveness is significant for getting both good naturalness and similarity of converted speeches.
Based on the finding, we propose a novel VC method utilizing deep hierarchical VAEs (DHVAEs) \cite{kingma2016biva,maaloe2019biva,larochelle2020nvae,child2021very}, which have high model expressiveness thanks to their hierarchical latent representations.
In addition, the conversion is relatively fast thanks to the absence of autogressive decoder which is often used in neural VC approaches \cite{van2017neural,liu2020vcc}.
However, DHVAEs cannot be used for VC in the same way as conventional VAE because they are unconditional models and have hierarchical latent variables.
Therefore, we propose a novel model called conditional deep hierarchical VAE (CDHVAE), which can perform VC by splitting the latent variables into speaker-dependent and invariant variables and inferring to the speaker-invariant latent variables.
Also, our RD analysis reveals that the mere use of high model expressiveness is insufficient because similarity can be degraded when the latent variable has redundant speaker information.
We address the problem by controlling the information contained in the latent variable using $\beta$-VAE objective.

The contributions of this paper are summarized as follows.
\begin{itemize}
    \item With an analysis from the RD perspective, we show that an increasing expressiveness of the model and $\beta$-VAE objective are significant for getting good naturalness and similarity in VAE-VC.
    \item We propose CDHVAE as one of the instances of VAEs with high model expressiveness. CDHVAE achieved mean opinion scores (MOSs) higher than 3.5 on both naturalness and similarity in inter-gender settings, which outperforms existing autoencoder-based VC methods.
\end{itemize}

\section{Preliminaries}

\subsection{Problem Statement}

In this paper, mel-spectrogram is used as an acoustic feature.
Let $x \in X$ be a segment of mel-spectrogram where $X$ is the ($80 \times T$)-dimensional Euclidean Space.
Here, $80$ is the dimension of the features.
$T$ is the sequence length of a segment, and we set $T=40$, which corresponds to 0.5 seconds, in our experiments.
Note that an utterance would be split into segments with a sequence length $T=40$ without overlapping at a conversion step.
Also, let $y \in Y$ be the speaker label where $Y$ is the group of all speakers.
The training set $D=\{ \{x_1, y_1\}, ..., \{x_m, y_m\} \}$ contains $m$ pairs of $(x_i, y_i) \in (X,Y)$, where $x_i$ is produced by the speaker $y_i$.
Given a tuple of a source speech, its speaker label, and target speaker label $(x_s, y_s, y_t)$, the goal of VC is to obtain an acoustic feature $x_{s \to t}$ that contains speaker characteristics of $y_t$ while preserving the linguistic content of $x_s$.

\subsection{VAE-VC} \label{sec:vae-vc}

Here we present a brief overview of VAE-VC \cite{hsu2016vaevc,hsu2017vawgan}.
VAE-VC approaches are aimed at obtaining speaker-invariant latent variables that contain only linguistic information by reconstructing speech from the latent variables and speaker labels.
Specifically, the methods consider a data generating process where an acoustic feature $x$ is generated from the latent variable $z$ and speaker label $y$, and parameterize the process with a deep neural network (DNN) decoder $p(x|z, y)$.
Here, $z$ and $y$ are defined to be independent, i.e., $p(z, y)=p(z)p(y)$.
The independence encourages $z$ to model information invariant to the speaker, i.e., linguistic information.
The objective of the VAE is given as follows:
\begin{align}
  \max \mathcal{L} =& - D_{KL} [q(z|x, y)||p(z)] + \mathbb{E}_{q(z|x, y)}[\log p(x|z, y) ] \nonumber \\
           \eqqcolon& - \mathrm{KL} - \mathrm{Rec}, \label{eq:cvae}
\end{align}
where the first term is the Kullback-Leibler (KL) divergence between the encoder $q(z|x, y)$ and the prior $p(z)$, and the second term is a negative reconstruction error.

Using the trained VAE, the conversion step is performed by decoding the speech from the speaker-invariant latent variable and the label of the target speaker.
Specifically, first, the encoder $q(z_s|x_s, y_s)$ is used to extract speaker-invariant latent variables $z_s$ from the source speech $x_s$.
Then, VC can be performed by inputting $z_s$ and target speaker $y_t$ to the decoder $p(x | z_s, y_t)$.

\subsection{Deep Hierarchical Variational autoencoders} \label{sec:dhvae}

DHVAEs, which are also called as bidirectional-inference autoencoder \cite{lee2021nvaetts}, are the family of VAE variants with high model expressiveness \cite{kingma2016biva,maaloe2019biva,larochelle2020nvae,child2021very}.
They share the same data generating process while their architectural details are different for each study.
It has three components parameterized by DNNs: an encoder $\Pi_{l=1}^L q(z_{l}|x, z_{<l})$, prior $\Pi_{l=1}^L p(z_l|z_{<l})$, and decoder $p(x|z)$.
Here, contrary to standard Gaussian prior for vanilla VAEs, the prior of DHVAEs enhances the expressiveness by hierarchically modeling $L$ latent variables $z = \{z_1, ..., z_{L}\}$, where $L$ is a hyperparameter.
The objective is given as follows:
\begin{align}
  \max \mathcal{V} =& - \sum_{l=1}^L \mathbb{E}_{q(z_{<l}| x)} D_{KL} [q(z_l|x, z_{<l})||p(z_l|z_{<l})] \nonumber \\
                    & + \mathbb{E}_{q(z|x)}[\log p(x|z) ], \label{eq:nvae} 
\end{align}
which also consists of the KL and reconstruction terms.

\section{Proposed Approach}

\subsection{Rate-distortion perspective on VAE-VC} \label{sec:tradeoff}

\begin{figure}[t]
  \centering
  \includegraphics[width=0.40\textwidth]{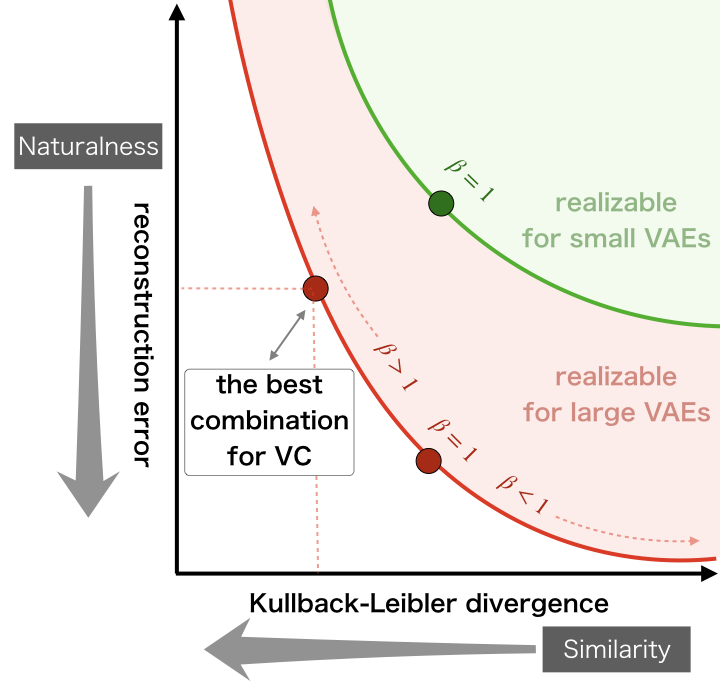}
  \caption{Schematic representation of the rate-distortion analysis on VAE-VC.
           y-axis measures the reconstruction error, which affects naturalness.
           x-axis measures the $\mathrm{KL}$ divergence between the encoder and the prior, which affects similarity.
  }
  \label{image:RD} 
\end{figure}

\begin{figure*}[t]
  \centering
  \includegraphics[width=0.85\textwidth]{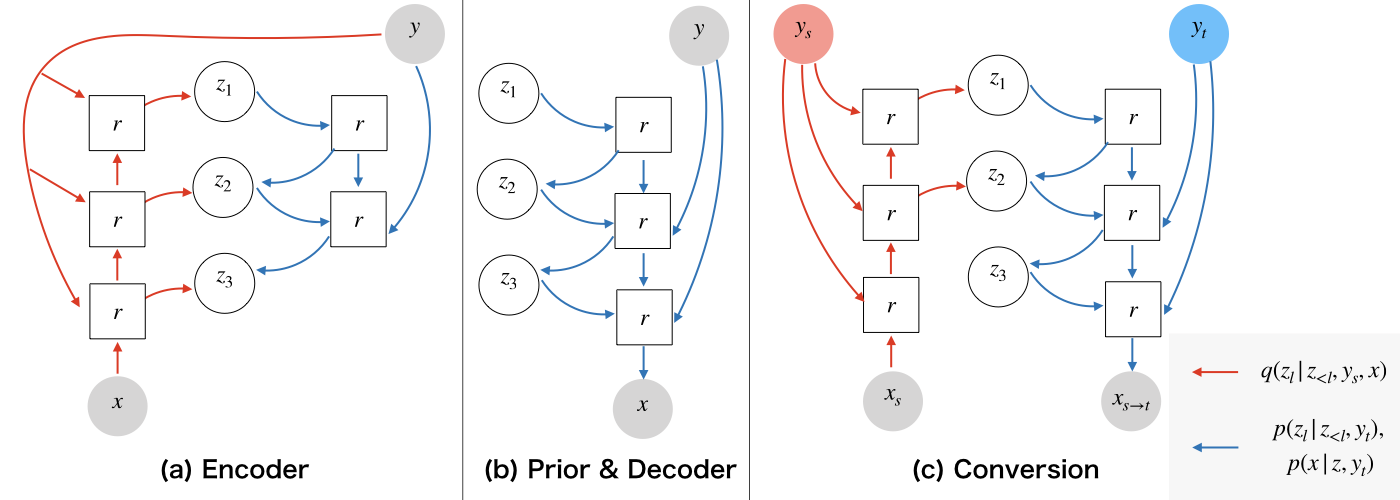}
  \caption{Schematic diagrams of CDHVAE with $L \coloneqq 3$ and $K \coloneqq 2$.
  (a) Inference process of the encoder $\Pi_{l=1}^{L} q(z_l|x, y, z_{<l})$.
  (b) Generating process of the decoder $p(x|z, y)$ and the prior $\Pi_{l \leq K} p(z_l|z_{<l}) \Pi_{l>K}p(z_l|z_{<l}, y)$.
  (c) Conversion process. Note that there is no red arrow from $r$ to $z_3$ in (c) because $z_3$ is sampled from prior at conversion step.
  Here, $r$ denotes residual unit that supports hierarchical modeling (further information can be found in \cite{kingma2016biva,larochelle2020nvae}).
  }
  \label{image:model} 
\end{figure*}

In this section, we adapt the RD analysis on VAE from \cite{alemi2018fixing} to voice conversion setting.
Based on the analysis, we will point out that it is necessary to use VAE with high model expressiveness for getting both good naturalness and similarity.
Speech naturalness and similarity are subjective measures. 
The former qualifies how natural (human-like) the converted speech $x_{s \to t}$ is.
The latter qualifies how similar the converted speech $x_{s \to t}$ is to that of the target speaker $y_t$.
Here, the {\em rate} equals to the KL divergence term ($\mathrm{KL}$), which measures the degree of compression.
Also, the {\em distortion} equals to the reconstruction error term ($\mathrm{Rec}$).

The RD analysis givin in \cite{alemi2018fixing} suggests that there is a trade-off between $\mathrm{KL}$ and $\mathrm{Rec}$, given a model architecture and the accompanied expressiveness.
Specifically, the phase diagram in Figure \ref{image:RD} shows the realizable combination of $\mathrm{KL}$ and $\mathrm{Rec}$ for a given model (either large or small).
Also, the Pareto frontier, i.e., the optimal combination of $\mathrm{KL}$ and $\mathrm{Rec}$, is called a RD curve.
The RD curve consists of a set of models trained by the following {\em $\beta$-VAE objective}, in which the weighting parameter $\beta \geq 0 $ is introduced into Eq. \ref{eq:cvae}:
\begin{align}
  \max \mathcal{L}_\beta =& - \beta \mathrm{KL} - \mathrm{Rec}. \label{eq:beta-vae}
\end{align}
When $\beta=1$, the objective targets a single point on the RD curve with slope 1 because $\mathrm{Rec}$ (y-axis) can be expressed as a linear function $\mathrm{Rec} = - \mathcal{L} - \mathrm{KL}$.
Setting $\beta < 1$ leads the situation where $\mathrm{Rec}$ is low but $\mathrm{KL}$ is high while $\beta > 1$ leads the opposite case (high $\mathrm{Rec}$ and low $\mathrm{KL}$), all without having to change the model architecture.

Next, we summarize the relationship between the RD curve and VC performance.
(i) First, low $\mathrm{Rec}$ is a necessary condition for high naturalness.
This is because high $\mathrm{Rec}$ disables the decoder to construct natural speech, and degrades the naturalness of the converted speech sampled from the decoder.
(ii) Second, low $\mathrm{KL}$ is a necessary condition for high similarity.
This is because high $\mathrm{KL}$ suggests that $z$ contains much information about $x$, which contradicts the intention of obtaining speaker-invariant latent variables.
On the other hand, low $\mathrm{KL}$ indicates that $z$ is well-compressed so redundant speaker information is lost from $z$ (note that the speaker information is inherently redundant for $z$ because the decoder is conditioned on $y$).
Based on (i) and (ii), we conclude that $\mathrm{Rec}$ and $\mathrm{KL}$ affect naturalness and similarity, respectively.
Note that, however, low $\mathrm{KL}$ and low $\mathrm{Rec}$ are necessary conditions but not sufficient on their own: to obtain high naturalness and similarity both 
$\mathrm{KL}$ and $\mathrm{Rec}$ need to be low.
For example, even if $\mathrm{Rec}$ is low, the naturalness can be low when $\mathrm{KL}$ is so large that the source and target speaker information is entangled in the inferred latent variables.

Based on the analysis, we can take two approaches to getting good naturalness and similarity.
One is using a hyperparameter $\beta$ in Eq. \ref{eq:beta-vae} to achieve the best combination of $\mathrm{KL}$ and $\mathrm{Rec}$ for a given model (note that there is no guarantee that $\beta=1$ achieves the best combination for VC).
The other is using a VAE with higher model expressivess to expand the realizable combinations, i.e., to move the RD curve to the left.

\subsection{Conditional Deep Hierarchical VAE}

Based on the analysis in Section \ref{sec:tradeoff}, we propose using a VAE with high model expressiveness, as well as using the $\beta$-VAE objective.
Namely, this paper adapts DHVAEs (see, Section \ref{sec:dhvae}) to VC tasks because they have high model expressiveness thanks to the hierarchical architecture.
However, they cannot be naively applied to VC tasks because they are unconditional models and have hierarchical latent variables.
Then, we propose CDHVAE, which separates speaker-dependent and invariant latent variables to perform VC.

First, we present a data generating process of CDHVAE, where the conditioning with speaker label $y$ is added to DHVAEs.
Namely, as well as DHVAE, CDHVAE has encoder, decoder, and prior.
However, the three components are conditioned with $y$ as follows:
\begin{align}
  \mathrm{Encoder} &= q(z_l|x, z_{<l}, y)  \;\; \forall l \leq L, \nonumber \\
  \mathrm{Decoder} &= p(x|z, y), \nonumber \\
  \mathrm{Prior}   &= 
      \begin{cases}
           p(z_l | z_{<l})        & \text{if}  \; l \leq K, \nonumber \\
           p(z_l | z_{<l}, y)     & \text{else}. \label{eq:conditioning}
      \end{cases}
\end{align}
The data generating process is illustrated in Figure \ref{image:model}-(a, b).
Here, $p(z_{\leq K})$ and $p(z_{>K}, y)$ are the speaker-invariant and dependent latent variables, respectively.
Also, $K$ is a hyperparameter that controls the expressiveness of the speaker-invariant and dependent latent variables.
It is because $p(z_{\leq K})$ becomes more expressive when $K$ becomes large, and vice versa ($p(z_{>K}, y)$ becomes more expressive when $K$ becomes small).
More specifically, as $K$ increases, the hierarchy of $p(z_{\leq K})$ increases, so the model expressiveness of $p(z_{\leq K})$ increases.
Then, a significantly small $K$ makes it difficult for speaker-invariant variables $z_{\leq K}$ to model linguistic information, which contradicts the intention of VAE-VC.
Also, a significantly large $K$ makes it difficult for speaker-dependent variables $z_{>K}$ to model speeches, which can degrades output speech quality.

Using the components, the objective of CDHVAE is given as follows:
\begin{align}
  \max &\mathcal{J}_\beta = - \beta \sum_{l=1}^K \mathbb{E}_{q(z_{<l}| x, y)} D_{KL} [q(z_l|x, z_{<l}, y)||p(z_l|z_{<l})] \nonumber \\
                          & - \beta \sum_{l=K+1}^L \mathbb{E}_{q(z_{<l}| x, y)} D_{KL} [q(z_l|x, z_{<l}, y)||p(z_l|z_{<l}, y)] \nonumber \\
                          & + \mathbb{E}_{q(z|x)}[\log p(x|z) ].
\end{align}
Here, we introduce a weighting parameter $\beta$ to balance the KL and reconstruction terms, as was done in Eq. \ref{eq:beta-vae}.

Using the trained CDHVAE, VC can be performed with the following procedure.
First, the speaker-invariant latent variable $ z_{\leq K}$ is obtained from the encoder $q(z_{\leq K} | x_s, y_s)$, given $x_s$ and $y_s$.
Then, the converted speech $x_{s \to t}$ can be obtained by inputting $z_{\leq K}$ and the target speaker label $y_t$ to the prior and the decoder.
This data generating process is illustrated in Figure \ref{image:model}-(c), and can be expressed as follows:
\begin{align}
  x_{s \to t} \sim \mathbb{E}_{q(z_{\leq K}|x_s, y_s)} \mathbb{E}_{\Pi_{l=K+1}^L p(z_{l} | z_{<l}, y_t)} p(x|z, y_t).
\end{align}
In practice, we approximate expectations with the mean of $q(z_{\leq K}|x_s, y_s)$ and  $\Pi_{l=K+1}^L p(z_{l} | z_{<l}, y_t)$.
Also, we split an utterance $x_s$ into segments with a sequence length $T=40$ and perform segment-wise conversion to adopt the same settings as training, but in principle it is possible to perform utterance-wise conversion.

The remaining challenge is what neural network architecture should be used to parameterize the data generating process of CDHVAE and to incorporate speaker labels.
We choose to adapt the model architecture of Nouveau VAE (NVAE) \cite{larochelle2020nvae} for CDHVAE, which is mainly composed of depthwise convolutions \cite{vanhoucke2014learning,chollet2017xception} and recently achieved state-of-the-art results in image generation.
However, because the original NVAE is an unconditional model, we add a simple modification for enabling conditioning: replacing all Batch Normalization (BN) layers with Conditional Instance Normalization (CIN) \cite{huang2017arbitrary,chou2019one}.
Specifically, the original NVAE have BN layers in components called residual cells which correspond to $r$ in Figure \ref{image:model}.
Then, we obtain speaker embedding from one-hot label $y$ using linear transformation, and use them as scale and location parameters in CIN.
Also, since NVAE has been proposed in the literature on image generation, the input should be ($C, H, W$)-dimensional tensor, where $C$ is the channel size, $H$ is the image height, and $W$ is the image width.
Therefore, we set $C=1$, $H=80$, and $W=T=40$ such that the mel-spectrogram can be used instead of an image.

\section{Related Works} \label{sec:related}

There are two VC approaches that utilize pure unparallel corpus and require neither parallel corpus nor text transcription: generative adversarial network (GAN) \cite{kaneko2018cycle,kameoka2018stargan} and autoencoder-based methods including VAE-VC.
VAE-VC was initially proposed by \cite{hsu2016vaevc}.
Subsequent research proposed auxiliary tasks to improve the naturalness \cite{hsu2017vawgan} and similarity \cite{kameoka2018acvae,tobing2019non} of the converted speech.
Also, some studies utilized vanilla autoencoder with some techniques (adversarial training \cite{chou2018multi} or architectural constraint \cite{qian2019autovc}) instead of VAE.
These methods are common to VAE-VC in that they attempted to obtain speaker-invariant latent variables;
therefore, these methods also may benefit from increasing model expressiveness as discussed in Section \ref{sec:tradeoff}.
In addition, similar to our method, some VAE-VC methods \cite{ho2020non,wu2020vqvc+} adopt U-Net-like architectures \cite{ronneberger2015u} and create hierarchical latent embeddings.
However, our method is different in that it adopts a very large hierarchy (e.g., $L=35$ in our experiments);
moreover, we show that the performance of VC significantly benefits from increasing model expressiveness, even without vector quantization which they employed.

From a technical perspective, our study adapts DHVAEs for voice conversion.
In the literature on image generation, DHVAEs recently achieved competitive performance compared to other deep generative models (e.g., GANs and autoregressive models) \cite{larochelle2020nvae,child2021very}.
While these DHVAEs are very recently applied to text-to-speech (TTS) \cite{lee2021nvaetts,liu2021vara}, they have not been applied to VC so far.
In addition, applying DHVAEs to VC requires unique techniques such as performing inference to speaker-invariant latent variables and controlling the balance of rate and distortion.

\section{Experiment}

\subsection{Settings}

In this section, we will evaluate CDHVAE on many-to-many VC tasks.
We strongly encourage readers to listen to the demos that can be found in \url{https://dena.github.io/CDHVAE/}.
The evaluation is performed on the VCTK corpus \cite{veaux2016superseded}.
While the original corpus includes utterances from 109 speakers, we trained the models on 20 speakers (10 females and 10 males), as performed in \cite{chou2018multi,qian2019autovc}.
The acoustic features were mel-spectrograms extracted from 48kHz audio.
As the vocoder, we used MelGAN \cite{kumar2019melgan}, which was trained on the same corpus.

CDHVAE is trained with a batch size of 8 for 200 epochs.
Regarding the model architecture, we adapt {\em CelebA model} in Table 6 of \cite{larochelle2020nvae} with two modifications:
replacing BN with CIN for conditioning, and removing normalizing flows for faster training.
The other training settings are the same with the official NVAE implementation \footnote{\url{https://github.com/NVlabs/NVAE}}.
In the CelebA model, the number of latent variables is set to $L=35$.
Also, based on our informal preliminary experiments, the hyperparameter in Eq. \ref{eq:conditioning} is set to $K=10$.
While this paper does not include hyperparameter study regarding $K$ due to computational resource limitation, we recommend not to use very small values nor very large values since $K$ balances the expressiveness of the speaker-invariant and speaker-dependent latent variables.
CDHVAE is compared to the recent autoencoder-based method proposed by \cite{chou2018multi}, whose pretrained-model is publicly available \footnote{\url{https://github.com/jjery2243542/voice_conversion}}.
Also, CDHVAE was trained with various values of $\beta \in \{1, 10, 50\}$.
Here, because we observed that CDHVAE with $\beta = 1$ resulted in high naturalness but low similarity in our preliminary experiments, we did not test $\beta < 1$.

We conducted naturalness and similarity tests using MOSs in the same manner as in previous studies \cite{qian2019autovc,yi2020vcc}.
We first selected one sentence for each of randomly selected 10 speakers (5 males and 5 females), and converted it to one randomly selected male speaker and one randomly selected female speaker, resulting in 20 utterances composed of 5 male-to-male (M2M), 5 male-to-female (M2F), 5 F2M, and 5 F2F samples.
Then, subjects in Amazon Mechanical Turk were asked to evaluate 80 utterances (20-utterance times 4-method).
After applying post-screening \cite{protasioribeiro2011crowdmos}, the answers for 29 and 45 subjects, who rated more than 70 utterances, were collected for naturalness and similarity tests, respectively.

\subsection{Results}


\begin{figure}[t]
  \centering
  \includegraphics[width=0.45\textwidth]{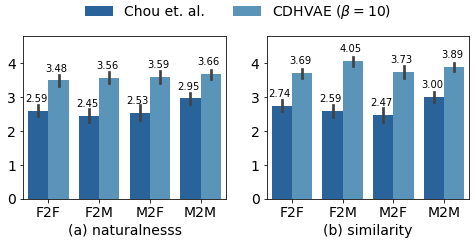}
  \caption{Comparing the baseline and CDHVAE using MOS. The error bars show 95\% confidence intervals.
  }
  \label{image:exp1} 
\end{figure}

From the MOS results, we can make the following observations.
{\bf (i) Using the VAE with high model expressiveness improved naturalness and similarity:}
the results in Figure \ref{image:exp1} shows that CDHVAE with $\beta=10$ achieved higher performance than the baseline method \cite{chou2018multi}.
Moreover, it achieved MOSs higher than 3.5 on both naturalness and similarity in inter-gender settings.
To give readers a better idea of what this means, notice that none of the existing autoencoder-based VC methods (e.g., AutoVC \cite{qian2019autovc} nor the methods in VCC 2020) reached 3.5 on naturalness and similarity at the same time.
Even though these are not fair comparisons due to the difference in experimental settings (e.g., training data or Hz of speeches),
it is encouraging to see that even with the simple $\beta$-VAE objective, the increasing expressiveness of VAEs enabled competitive or higher quality compared to the existing autoencoder-based VC methods.

\begin{figure}[t]
  \centering
  \includegraphics[width=0.45\textwidth]{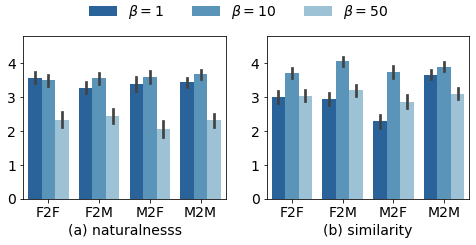}
  \caption{Hyperparameter study on $\beta$ using MOS. The error bars show 95\% confidence intervals.
  }
  \label{image:exp2} 
\end{figure}

{\bf (ii) $\beta$ is an important factor for balancing the tradeoff between naturalness and similarity:}
the results in Figure \ref{image:exp2} shows that CDHVAE with low $\beta$ value ($\beta=1$) achieved lower similarity scores than those with $\beta=50$ in inter-gender settings.
This indicates that a small $\mathrm{KL}$ value is a necessary condition for good similarity.
On the other hand, CDHVAE with high $\beta$ value ($\beta=50$) achieved the lowest naturalness scores because the very large $\beta$ makes reconstruction difficult.
This indicates that a small $\mathrm{Rec}$ value is a necessary condition for good naturalness.
Here, note that small $\mathrm{KL}$ and $\mathrm{Rec}$ values are merely the necessary conditions as noted in Section \ref{sec:tradeoff}, and properly selected $\beta=10$ achieved the highest MOSs for both naturalness and similarity.

\textbf{Conversion speed:}
On a 16-GB Tesla T4 GPU, we can convert a segment of mel-spectrogram of the size 80 $\times$ 40 in 0.172 seconds (344 ms / 1-second speech).

\section{Discussions}

In the experiment using VCTK corpus, the proposed model called CDHVAE achieved MOSs higher than 3.5 on both naturalness and similarity in inter-gender settings.
While we used simple $\beta$-VAE objective for CDHVAE, the performance could be improved by combining with the existing auxiliary losses for VAE-VC.
Other future work may be to train CDHVAE with expressive or noisy speech corpora because the possible merit of VAE-VC is not requiring a pre-trained speech synthesis model or mapping from PPG to speech.

\section{Acknowledgements}
This work was supported by JSPS KAKENHI Grant Number JP20J11448.

\bibliographystyle{unsrt}
\bibliography{mybib}
\end{document}